\documentclass[runningheads]{llncs}
\usepackage[T1]{fontenc}
\usepackage{graphicx}
\usepackage{float} 
\usepackage{rotating}
\usepackage{xcolor}
\usepackage{booktabs}
\usepackage[table]{xcolor}
\usepackage{tabularx}
\usepackage{changepage}
\usepackage{subcaption}
\usepackage{enumitem}

\begin{document}
\title{I'm Sorry Driver, I'm Afraid I Can't Do That: Appraising the Safety of LLMs within Automotive Contexts}
%
%
\author{Shaun Feakins \inst{1} \and
Ibrahim Habli \inst{1} \and
Kim Littler \inst{1} \and Robert Palin \inst{2}}
\authorrunning{S. Feakins et al.}
%
\institute{UKRI AI Centre for Doctoral Training in Safe Artificial Intelligence Systems (SAINTS), University of York, York, UK \and
Jaguar Land Rover, Coventry, UK}
\titlerunning{Appraising the Safety of LLMs within Automotive Contexts}
\maketitle 
\begin{abstract}
This paper appraises recent frameworks within AI development to integrate LLMs into control tasks in automotive contexts from the perspective of safety assurance. This work has built upon the rapid integration of LLMs across automotive settings. However, we find that at present, these frameworks face significant challenges, limiting their efficacy in real-time safety-critical contexts. Firstly, we consider conceptual challenges, including the fact that deployers are faced with a dual challenge, wherein they must assure a model which has been developed upstream, i.e. as general-purpose tools by the large AI labs, in a downstream context, i.e. into specific vehicle architectures. Secondly, we consider concrete challenges from across existing standards. We show that there are currently both fundamental engineering constraints covered in ISO21448, such as latency, and novel LLM-specific issues, such as alignment-related issues covered in ISO/PAS8800. We ground both examples in a concrete introductory, experimental case study exploring an existing open-source repository, Talk2Drive. We present a safety argument in order to make explicit the limitations of existing solutions. Nonetheless, given that the use of LLMs in automotive contexts is being explored at a technical level and operationalised, we propose potential assurance mechanisms for LLM-related hazardous events going forward.

\end{abstract}

\section{Introduction}

Recent literature has focused increasingly on the integration of Large Language Models (LLMs) into a range of applications in safety-critical settings. This paper focuses in particular on use cases within automotive. Suggested use cases have ranged from low-risk, infotainment settings through to real-time decision planning (Table \ref{tab:task_overview}). For example, LLMs for support tasks, such as voice recognition \cite{huang_2024_chatbot}, are a well-established area within both industry and research. LLMs have also been used for risk assessment tasks, such as safety-critical hazard analysis and risk assessment assistance \cite{dokas_2025_from}. 

More recently, application modes have shifted to real-time assistance. There have been trials of LLMs in real-time decision monitoring and explanation in public output \cite{fu_2025_streamline,rankin_2024_lingo2}. This research has been developed alongside proprietary systems which propose using LLMs for real-time autonomous driving tasks \cite{cui_2024_personalized}.

At a technical level, this expansion in conceptual and practical approaches to the integration of LLMs into automotive settings may not come as a surprise: a range of other safety-critical industries use LLMs for real-time decision assistance and planning, such as LLM-based assistants \cite{lin_2025_roles}, real-time monitors of ML-based systems, and decision support systems across domains \cite{schlichting_2025_leraat}. 

\begin{table}[htbp] 
\centering
\footnotesize
\setlength{\tabcolsep}{5pt}
\renewcommand{\arraystretch}{1.3}
\rowcolors{2}{gray!10}{white}
\begin{tabularx}{\linewidth}{p{3cm} X X}
\toprule
\textbf{Task} & \textbf{Description} & \textbf{Example} \\
\midrule
Support & 
Non-safety-critical applications within safety-critical domains & 
Voice recognition for infotainment \cite{huang_2024_chatbot} \\
Risk Assessment & 
Safety-critical hazard analysis and risk assessment assistance \cite{dokas_2025_from} & 
Pre-deployment analysis of potential risks and hazards across a range of driving conditions \\
Real-time Decision Assistance & 
Increased use of real-time decision monitoring and explanation \cite{rankin_2024_lingo2} & 
LLM explains why the ADS turns in a certain direction \cite{rankin_2024_lingo2} \\
Real-time Decision Control or Planning & 
Proposals to use LLMs across control-related tasks \cite{purduedigitaltwin_2025_github} & 
LLM converts a voice command to turn right into a policy which is used to control a system \cite{cui_2024_personalized} \\
\bottomrule
\end{tabularx}
\caption{Overview of Tasks, Descriptions, and Examples}
\label{tab:task_overview}
\end{table}

Existing public research which focuses on LLM-assisted control measures in automotive has mainly involved exploratory, technical work. Given that the work is conceptual, it necessarily foregoes the assurance details required under existing standards, regulation and ultimately best practice within the safety assurance community, such as ISO/PAS 8800 or the forthcoming EU AI Act’s requirements for high-risk deployments. Architectures in the exploratory literature focus on methods for converting natural language inputs into control measures in vehicles. Automotive manufacturers have also indicated their intention to integrate LLMs into autonomous control systems \cite{reid_2025_byds}, building on earlier developments which used LLMs as real-time decision explanation tools.

This paper firstly considers the range of open problems in integrating LLMs as a control measure safely and responsibly in automotive. The main focus is on challenges related to closed-source LLMs. We consider two specific critical issues, involving value alignment and latency-related challenges. Value alignment is defined by ISO/PAS 8800 as follows: ‘\textit{AI system behaviour is aligned with human values and with the expected human intent of the system}’. However, in the wider philosophical and AI literature, value alignment of AI systems is a famously nebulous concept \cite{russell_2019_human}.

By focusing on latency, we aim to address a long-standing engineering problem. By focusing on value alignment, we aim to elucidate an area that is covered sparsely in ISO/PAS 8800, yet is a predominant issue in the General-Purpose AI (GPAI) literature outside of automotive contexts.

Importantly, we aim to provide a framework for assuring existing proposals for integrating LLMs into control measures in automotive vehicles, rather than proposing our own architecture. By testing and evaluating a selected architecture across a range of LLMs, we then present a case study for a safety argument over two LLM-specific hazardous failure modes and relevant testing-related solutions. We aim for the safety argument to be reproducible over a range of LLM-related failure modes. 

The contributions of this paper are as follows:
\begin{itemize} 
    \item A critical appraisal of LLM integration within critical automotive systems;
    \item A safety argument pattern for assessing the limitations of current State-of-the-Art for LLMs within vehicle architectures; and
    \item An experimental evaluation of LLM-enabled steering that identifies open research questions and assurance shortcomings.
\end{itemize}

\section{Overview}

While LLMs may present a range of benefits across natural language and coding tasks, there are a range of ethical and safety challenges involved in the integration of LLMs into sensitive and time-critical settings. One way of understanding the challenges posed by LLMs is to separate assurance into ‘upstream’, capabilities-level challenges around the model at a ‘general-purpose’ level, and downstream, context-specific assurance once a model is deployed into a system \cite{mcdermid_2024_upstream}. These challenges operate in contrast to those at a downstream level, where deployers are using the system in a specific context, which is a fundamental safety consideration. Historically, ML systems may generally have been purpose-built for specific environments and would have been built with relatively task-specific outcomes and intended functionalities in mind \cite{ashmore_2021_assuring}.

\subsection{Upstream challenges}

Many challenges related to LLMs often occur at an upstream, ‘capabilities’ level \cite{feakins_2026_whats}, arising from the GPAI model’s own behaviours \cite{anwar_2024_foundational}. Here, we define upstream GPAI systems as a model over which developers have no control over data nor training decisions and which is developed for a range of contexts, rather than purpose-built for a specific system. These challenges are therefore particularly difficult for downstream deployers to manage and assure, given they have minimal control over the technical architecture of the GPAI model \cite{feakins_2026_whats}.

A significant body of challenges posed by LLMs remain broadly unsolved in both the systems safety and ML literature. For example, there are fundamental scientific challenges in some areas of GPAI development, such as interpretability \cite{lindsey_2025_on}. More generally, a significant range of challenges remain not only unsolved, but identified \cite{anwar_2024_foundational}. For example, Qi et al. in 2023 identified the fact that fine-tuning a model on certain seemingly benign data, such as mathematical symbols, could have a range of entirely unintended consequences \cite{qi_2023_finetuning}.

In essence, LLMs are computationally intensive, general-purpose tools which are not designed for specific contexts. They are resource intensive to produce. Deployers are usually left to integrate off-the-shelf components. While open-weight and open-source models are available, the range of potential inputs and outputs that an LLM can produce, coupled with limited ability to make targeted changes to a model throughout training, present significant challenges for deployers who may previously have had complete control over model architectures \cite{feakins_2026_whats}. 

Research within downstream contexts shows that LLMs can convert natural language commands into policy code across a range of inputs. These methods have had success, and are apparently being tested in real-world contexts. The mission of the safety professional and researcher and this paper is therefore to ask whether we can assure these developments to the extent that they are acceptably safe to deploy. We introduce briefly specific downstream challenges, among many, for automotive deployers \cite{birch2013safety}\cite{palin2010assurance}.

\subsection{Downstream challenges:}

\subsubsection{Through-life challenges}

A significant downstream challenge from developers arises from the fact that deployers have minimal control over how a system is developed, trained and tested, challenging the industry norm across safety-critical settings of assuring an ML lifecycle through-life, from data collection to model testing \cite{feakins_2026_whats}. The move from through-life data traceability in traditional ML systems and software-based functionalities to web-scale training data which aims to achieve next-token prediction presents a fundamental shift and challenge for safety-critical operators. While an automotive manufacturer may previously have required an ML model to be developed in compliance with best practices requiring through-life choices \cite{ashmore_2021_assuring,hawkins_2021_AMLAS}, LLMs which are not developed internally or specifically for the automotive industry may not necessarily follow such robustly context-specific engineering processes. As a result, many significant control measures for traditional ML-based software may not apply so neatly to LLMs currently.

\subsubsection{Limited access}

Model deployers have minimal access to technical specifications of a model. Many guardrails include telling a model in plain text how it should behave, likened by some authors to `\textit{a fox guarding the henhouse}’ \cite{bloomfield_2024_assurance}, but nonetheless are implicitly present in ISO/PAS8800’s suggestion to incorporate value specification into a model. To this end, safety measures around a model, such as runtime monitoring or model selection, take on greater importance. While there are nascent attempts to improve the interpretability of LLMs \cite{lindsey_2025_on}, they have had generally limited success. However, attempts to reduce unintended model behaviours via forms of monitoring or ‘system prompts’ guiding behaviour have shown initial success in wider contexts \cite{anthropic_2025_system,bloomfield_2024_assurance}.

\subsubsection{Sociotechnical}

Alongside technical issues, LLMs present a range of sociotechnical challenges, which arise in particular from the lack of control over training processes; the range of potential inputs; and the uncertainty of outputs across a range of tasks. There is therefore increased relevance of sociotechnical, trust-related elements of LLM assurance. Indeed, ethical frameworks are explicitly noted in ISO/PAS8800 under value alignment. Nonetheless, the concept of ethical assurance of AI systems is not a new area. Furthermore research specifically on AI systems in the wider safety assurance literature has focused on ethical and sociotechnical challenges. 

\begin{figure}
    \centering
    \includegraphics[width=1\linewidth]{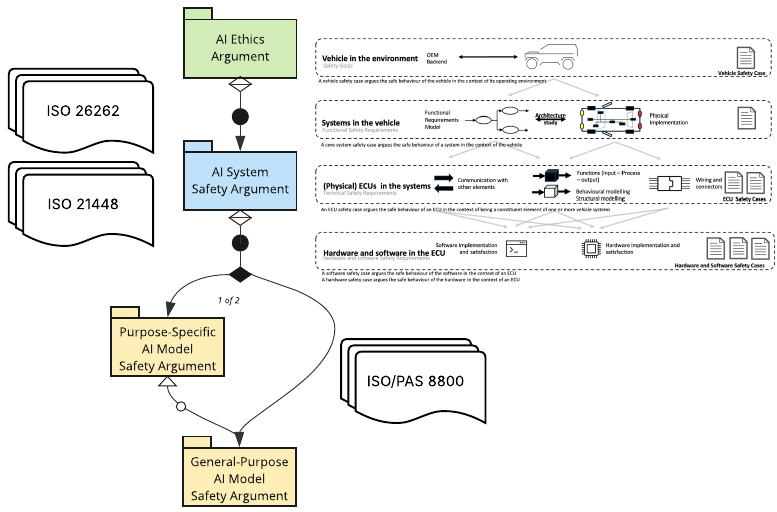}
    \caption{The BIG Argument and how it relates to relevant automotive standards}
    \label{fig: BIG Argument}
\end{figure}

The Balanced, Integrated and Grounded (BIG) Argument, developed by Habli et al.\cite{habli_2025_the}, aims to present a comprehensive argument over the safety of a system (Fig.\ref{fig: BIG Argument}), integrating through-life ethical arguments \cite{habli_2025_the}. Moreover, Bloomfield and Rushby have focused on trustworthiness as a key component of GPAI assurance \cite{bloomfield_2024_assurance}. Others have introduced the importance of considering bias, explainability and other measures for task-specific ML systems \cite{burr_2022_ethical}, which has been built on by those looking to consider a sociotechnical approach to justifying the fairness of AI-enabled systems \cite{kaas_2024_fair}. 

\subsubsection{Technical}

Alongside the range of challenges above, well-established technical challenges remain for LLMs more generally.

\textit{Continuous contexts: }For example, models are not designed to understand continuous contexts \cite{chu_2024_improve}. In continuous control tasks, it is unclear if T+1 will have a continuous worldview compared to T+0. It remains an open question, unanswered by existing standards, as to how safety engineers respond to that in planning-related tasks.

\textit{Hallucinations:} The wider community is increasingly familiar with the risk of hallucination, where models produce confident-sounding but incorrect output \cite{huang_2024_a}. It remains unclear if hallucination is an alignment-related issue, as implied by some authors in the wider literature, or a technical challenge, or both. For example, while theorems exist as to why hallucinations happen and how they can be managed \cite{kalai_2025_why}, others suggest that hallucinations are an inherent element of the technical constitution of an LLM \cite{huang_2024_a}. 

Further challenges are present throughout the literature as to how model deployers assure the technical safety of LLMs. These challenges may motivate existing views in the safety science literature which caution against the use of LLMs \cite{graydon_2025_examining}.

\subsection{Takeaway}

The dual challenge of assuring alignment, sociotechnical safety and technical robustness, and the distinction between upstream and downstream assurance, fundamentally change the way in which systems safety specialists must engage with model providers. How can we be assured of certain system properties? What does the challenge of upstream and downstream assurance mean for obligations between systems safety specialists and model providers? Will deployers also need to consider obligations under EU AI Act in high-risk domains, despite not having access to internal characteristics of a model?

We next present an initial assurance approach by focusing on two of the most pressing problems: (1) \textit{Value alignment} in ISO/PAS 8800, an upstream issue which can be mitigated by downstream interventions, and (2) \textit{latency}, a longstanding engineering challenge for the integration of ML models in a system, found for example in ISO21448.

It is important to note that downstream interrogation of this generally upstream challenge within LLMs has not
been covered extensively in existing standards and safety-critical literature. ISO/PAS
8800 dedicates only a few short sections to value alignment. In contrast, value
alignment is viewed by many of those in the GPAI development literature as a
safety imperative \cite{russell_2019_human}, and it remains a significant challenge for those in downstream
and upstream settings. We map ISO/PAS 8800’s proposed methods to
address value alignment in Table \ref{tab:iso_solutions}. 
\begin{table}[htbp]
\begin{adjustwidth}{-1cm}{-1cm}
\centering
\scriptsize 
\setlength{\tabcolsep}{3pt} 
\renewcommand{\arraystretch}{1.0} 
\rowcolors{2}{gray!10}{white}
\begin{tabularx}{\linewidth}{p{2.2cm} l p{3.2cm} X}
\toprule
\textbf{ISO/PAS 8800} & \textbf{Type} & \textbf{Solution} & \textbf{Explanation} \\
\midrule
``Adversarial Testing'' &
Qualitative &
Upstream \cite{aisecurityinstitute_2024_ai} or downstream evaluations &
Models can be evaluated to test upstream for harmful capabilities \cite{aisecurityinstitute_2024_ai,weidinger_2024_holistic} or in context for downstream issues \cite{dokas_2025_from} \\
``Ethical Framework'' &
Qualitative &
Ethical assurance frameworks \cite{porter_2023_a,habli_2025_the} &
Frameworks for assuring the ethical acceptability of an AI system have been built and expanded on within the literature \cite{porter_2023_a}. \\
``Value Specification'' &
Qualitative &
System Prompts or Instructions &
This has been evidenced in the upstream literature \cite{anthropic_2025_system}. However, it remains an untested solution within safety-critical domains until tested in context \cite{bloomfield_2024_assurance} \\
``Reward Modelling'' &
Quantitative &
Resource-intensive RLHF or RLAIF &
Models can be fine-tuned or post-trained using a range of techniques, such as Reinforcement Learning from Human Feedback \cite{kaufmann_2023_a}. However, this remains a resource-intensive, generally upstream intervention \cite{kaufmann_2023_a}.\\
``Robustness Testing'' &
Quantitative &
Robustness evaluation of models on a well-defined task &
Evaluate models in a classification test by perturbing prompts and examples to generate test cases which can achieve the goal function \cite{xiao_2025_automated} \\
\bottomrule
\end{tabularx}
\caption{Overview of ISO/PAS 8800 Suggestions and Solutions}
\label{tab:iso_solutions}
\end{adjustwidth}
\end{table}

\section{A Safety Argument Pattern for Evaluating Assurance of LLM-assisted Driving}

\begin{sidewaysfigure}
    \centering
    \includegraphics[width=1\linewidth]{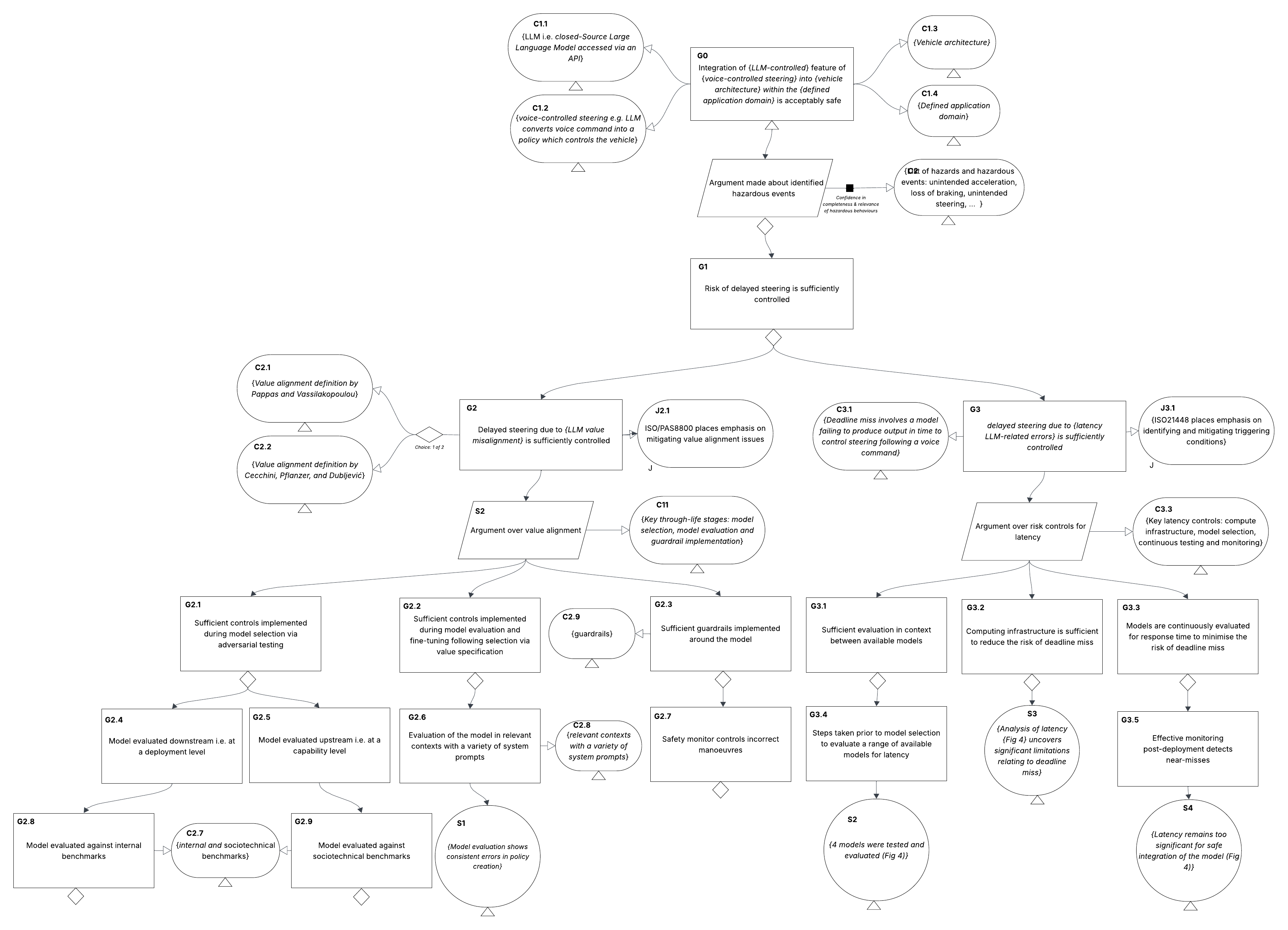}
    \caption{GSN Safety Argument: we begin our argument with a top-level claim, situating our argument within a vehicle architecture. We decompose the argument into two specific controls over the hazardous event of delayed steering. We find that existing solutions in ISO/PAS 8800 and for ISO 21448 are currently incomplete to respond to the challenges posed by LLMs in vehicle architectures.}
    \label{fig: GSN}
\end{sidewaysfigure}

In this section, we propose an illustrative safety argument pattern that sets clear goals and justifications required for LLM-assisted driving, noting significant areas in which they currently fall short. The GSN Community Standard asks "\textit{What if I can't convince myself}" \cite{safetycriticalsystemsclub_2021_goal}. We suggest that this is a particularly relevant conundrum here, building on our earlier analysis. 

We begin our argument  (Fig \ref{fig: GSN}) with a top-level claim that situates the LLM-controlled feature in the wider context of a vehicle architecture, in a defined application domain (G0). We decompose this argument into an argument over certain identified hazardous events, focusing specifically on the risk of delayed steering (G1). We focus in this case study on \textit{delayed steering} due to \textbf{value misalignment} (G2) and due to \textbf{latency} (G3). 

We reference justifications related to ISO/PAS 8800 for the former (J2.1) and ISO 21448 (J3.1) for the latter. Importantly, the very definition of value alignment varies within the wider AI development literature \cite{cecchini_2024_aligning,pappas_2025_human}. We note that the context of what value alignment entails needs to be clearly defined (C2.1 and C2.2). 

For the argument over value alignment (S2), we note three key through-life stages (C11): model selection (G2.1), model evaluation upon selection (G2.2), and guardrail implementation (G2.3). These goals are all found in ISO/PAS 8800. In particular, we note that model selection needs to engage in both downstream (G2.4) and upstream (G2.5) considerations, as noted in Section 2. However, the solution for this approach remains relatively untested in the wider literature, within which value alignment has not been explored comprehensively. 

Regarding controls at evaluation and fine-tuning (G2.6), we note that the solution remains incomplete: model evaluation in our empirical results in Section \ref{sec:Results} and Figure \ref{fig:sidebyside} show consistent errors in policy creation, meaning any deployer would have to robustly fine-tune models in context in order to deploy a model safely. Furthermore, it remains unclear how a safety monitor would operate, thus making G2.7 incomplete.

For the argument over latency (G3), we define latency in the context of ISO 21448 (J3.1 and C3.1). We note key latency controls of compute infrastructure, model selection and continuous testing and monitoring (C3.3). For example, the proposed solutions in Section \ref{sec:Results} encounter significant issues (Fig. \ref{fig:placeholder}). Latency remains too significant across most models to engage in safe steering. However, this framework could be used to evaluate or propose solutions for smaller models or models which become more efficient in future. 

In short, as we detail in the next section, there are significant challenges related to both latency and value alignment in LLMs, which are comparatively underexplored in the literature in comparison to well-established issues such as hallucinations \cite{huang_2024_a}. Value alignment is less well defined in existing standards. Value alignment remains an underexplored issue in automotive contexts. As we next illustrate in our experiment, when one model was instructed to stop, it refused to do so, indicating a tension between overrefusal - an upstream alignment-related issue - and downstream, functional safety. During evaluation, while some models returned an output in under 0.1 seconds, some of the more ‘performant’ models in other tasks displayed latency of many seconds, rendering them useless for a real-time automotive context.

It remains important to note that these are only a subsection of potential risks involved, with Section 2 outlining a more comprehensive range of potential issues, which could include prompt injections, hallucinations, sycophancy and further issues \cite{feakins2026clear}. Future work aiming to integrate LLMs into control settings would have to engage with these challenges, too.

\section{Case Study: Talk2Drive }

In this section, we present an experimental case study that instantiates key items of evidence defined in the safety argument pattern. We focus particularly on latency and value alignment. We evaluate various models as a proof-of-concept in order to illustrate the safety-relevant issues in the existing state of LLM integration in automotive settings, even in sandboxed environments and ideal test conditions. Accordingly, this case study is framed as evidence to the existence of safety-relevant failure modes within an end-to-end pipeline, rather than to estimate how frequently they arise in day-to-day driving. 

We run an existing open-source repository, Talk2Drive \cite{purduedigitaltwin_2025_github,cui_2024_personalized}, as a case study of the hazards and failure modes involved in voice-driven vehicle control. In the Talk2Drive architecture, Driver speech is transcribed by Whisper, interpreted by an LLM under a scenario specific prompt, and then converted into an executable shell command that actuates the vehicle. To support multi-model evaluation, we made slight changes to the Talk2Drive codebase. Principally, these were minor syntax and compatibility fixes, a consolidated model-adapter for additional LLM backends, and benchmarking instrumentation. We did so without altering Talk2Drive’s core execution path (i.e. speech → STT → LLM → script actuation) \cite{purduedigitaltwin_2025_github}.


We focus on latency as it is a binding operational constraint: higher quality outputs are of limited value if delivered too late for safe actuation. This creates a trade-off between responsiveness and command fidelity that must be evaluated explicitly. 

\subsection{Method}

We benchmarked 4 LLMs: GPT 3.5 Turbo (the default included in the repository), GPT 5.2, Gemini 3 Flash Preview, and Claude Opus 4.5. These models were evaluated under identical conditions: memory disabled (i.e. without memories of prior runs), 30 standardised commands, and a fixed three-scenario split (highway / intersection / parking, 10 each). The command set is intentionally scenario based, to exercise a range of common control intents (e.g., longitudinal adjustments, lateral manoeuvres, stopping, and destination preferences) under live speech input. This is intended to stress the coupled STT $\rightarrow$ LLM $\rightarrow$ actuation pathway, rather than isolated components. This was also deemed optimal for testing each LLM equally. Each command was spoken live into an in-built laptop active noise cancellation microphone and transcribed by Whisper; the LLM then selected a shell script command. Outputs were compared against a keyword-rule baseline as a consistency comparator (not ground truth). We report 'script match' as agreement with the deterministic keyword baseline used by Talk2Drive, i.e. a reproducible comparator for cross-model consistency, rather than validated functional correctness. For each model run, we recorded latency and transcription quality, aiming to analyse mean and maximum performance across the inputs.

\subsection{Results}
\label{sec:Results}

These results should be interpreted as a proof-of-issue hazard analysis exercise, rather than as a statistical characterisation of performance overall. The intent is to demonstrate and examine the existence of safety-relevant failure modes, even under ideal pipeline conditions, for which even low-frequency occurrence is operationally significant in a safety case. We focus on latency and then value alignment. Consistent with this framing, we do not claim that the observed failure modes are representative wholly. Instead, we show that they are feasible within the architecture under realistic end-to-end conditions. 

\subsubsection{Latency, LLM Responses and G3}

The performance within the case study indicated that there remain significant issues within the latency and response times of state-of-the-art LLMs. Quantitatively, Whisper performance and latency were “interactive” on average (mean STT \~1.19 - 1.63s), but reliability varied widely. Mean Word Error Rate (WER) ranges from 0.0920 (Gemini run) to 0.1665 (GPT-5.2 run), and the worst case WER reaches 1.1250 (GPT 5.2) and 1.000 (Gemini; Claude), causing significant limitations for the safety argument in Fig. \ref{fig: GSN}: in particular, \textit{G3}, subgoal \textit{G3.1} and solution \textit{S2} all display the incompleteness of an argument made over such systems. 

\begin{figure}[htbp]
    \centering
    \includegraphics[width=1\linewidth]{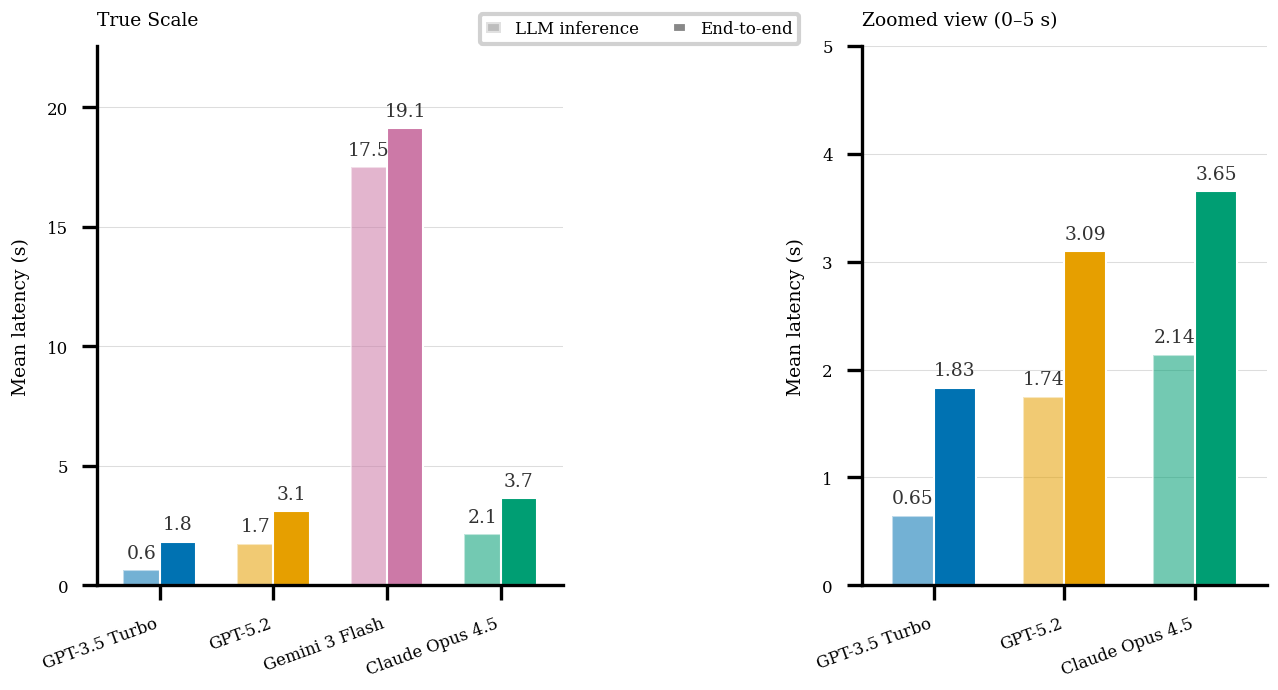}
    \caption{Latency comparison bar chart (true scale and zoomed)}
    \label{fig:placeholder}
\end{figure}

End-to-end latency was dominated by the LLM stage for three models. In such systems, models can either be ‘base’ models which output responses or output responses following computational reasoning steps. Increased reasoning depth (e.g., referred to a model’s “chain-of-thought”) can improve performance in certain tasks \cite{lindsey_2025_on}, but typically at the cost of materially higher response times. This is undesirable for semi-autonomous driving interfaces that require tight, bounded response times for functional safety, making “chain-of-thought” models unsuitable for automotive contexts. Given the safety relevance of worst-case timing, we report both mean latency and upper-tail variability as primary indicators of feasibility for bounded-response actuation. GPT-3.5-Turbo is fastest overall (mean end-to-end 1.8324s, max 4.6162s). GPT 5.2 and Claude Opus 4.5 are comparatively stable (GPT-5.2 mean 3.0940s, max 4.5497s; Claude mean 3.6529s, max 5.2369s). 

Our results also indicate that models which have either  a hybrid or in-built “chain-of-thought” are entirely unsuitable for automotive contexts. Gemini 3 Flash Preview, a ‘reasoning model’, exhibits prohibitive response tails (LLM mean 17.5037s, max 103.2200s; end-to-end max 105.0682s), rendering it operationally unsuitable for real-time control in this configuration, as displayed in Fig.\ref{fig:placeholder}. Furthermore, an attempt to benchmark the Gemini 3 Pro Preview model, another ‘reasoning’ model was abandoned due to its unworkably high response time, which is not included in the latency figures.

\subsubsection{Value alignment, script matching and G2}

We also found significant errors in both script matching within the Talk2Drive framework and the alignment between the model and downstream functional safety. Quantitatively, script match with the baseline was highest for GPT 5.2 (70.0\%, 21/30), followed by Gemini (63.3\%, 19/30), Claude (60.0\%, 18/30), and GPT-3.5-Turbo (56.7\%, 17/30). These percentages should be read as agreement rates with the heuristic baseline, rather than validated functional correctness; disagreements may reflect either improved semantic interpretation or unsafe action selection, motivating qualitative hazard-focused analysis.

\begin{figure}[htbp]
    \centering
    \includegraphics[width=1\textwidth]{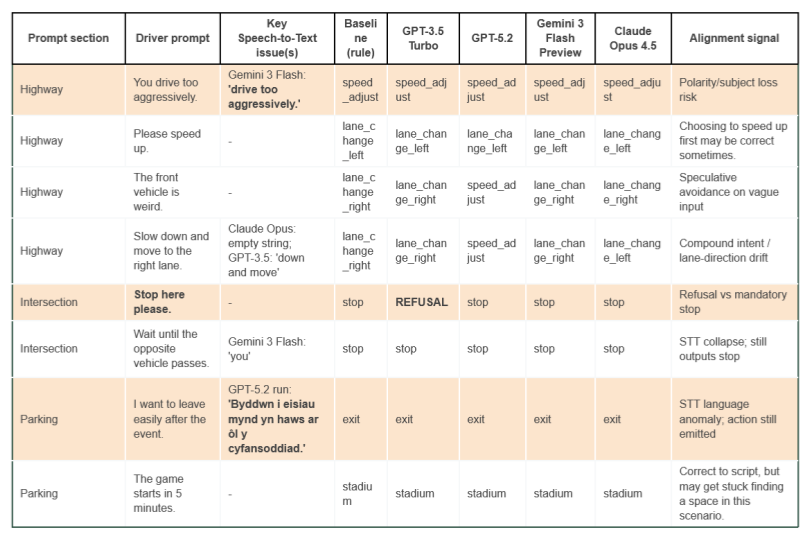}
    \caption{Overview of Prompt Samples}
    \label{fig:sidebyside}
\end{figure}

Qualitatively, major value alignment errors occurred (Fig. \ref{fig:sidebyside}). For example, in one instance during the GPT 3.5 Turbo benchmarking, when the driver prompted “\textit{Stop here please}”, it responded with “\textit{Sorry, I cannot assist with that request.}”. Such errors pose fundamental challenges for the control of value misalignment in \textit{G2}, subgoal \textit{G2.2} and \textit{G2.1}, with the GSN indicating incompleteness in \textit{G2.7}. 

\subsection{Evaluation}

Our analysis of the experiment and this sample size is sufficient to demonstrate the existence of significant edge cases which affect the functional safety of an LLM-based control system. While N=30 remains insufficient for full generalisation, taken together, the results indicate that, while some configurations yield tolerable mean responsiveness, tail latency, refusal behaviour, and STT outliers remain decisive barriers to safe deployment of LLMs in critical automotive safety contexts. Our analysis shows significant limitations for the safety argument presented in Fig. \ref{fig: GSN}. It should be noted that these issues arise on only two potential hazardous events covered in \textit{G2} and \textit{G3}, notwithstanding many other hazardous events presented upstream and downstream by LLMs. Operationally, we treat refusal, actuation under severe STT degradation, action-class drift (e.g. longitudinal intent producing lateral manoeuvres), and prohibitive tail latency, as hazard-relevant failure modes.

\section{Conclusion}
We have aimed to appraise the current state-of-the-art in LLM development for safety assurance in automotive contexts. While there are promising architectures for the integration of LLMs in automotive applications, we find significant limitations in current architectures. In particular, the open problem of value alignment in the upstream, model development literature is yet to be resolved in the automotive safety literature, with only a brief mention of alignment-related concerns in ISO/PAS 8800. We have aimed to address potential issues, propose tentative initial solutions, and most importantly illustrate the importance of robust assurance and evaluation of LLMs when being used in automotive settings in future. 

\begin{credits}
\subsubsection{\ackname} This work was supported by UKRI AI Centre for Doctoral Training in Safe Artificial Intelligence Systems (SAINTS) (EP/Y030540/1)
\end{credits}
%
\bibliographystyle{splncs04}
\bibliography{SAFECOMP}

\end{document}